\documentstyle[preprint,revtex]{aps}

\tightenlines

\begin{document}

\preprint{TRIUMF TRI-PP-92-99}
\preprint{\ }
\preprint{McGill 92-52}
\preprint{\ }
\preprint{October 1992}

\draft

\begin{title}
Quark model calculation of $\eta \to l^+ l^-$ \\
to all orders in the bound state relative momentum
\end{title}

\author{Bernard Margolis$^{(a)}$, John Ng$^{(b)}$,
Martin Phipps$^{(a)}$, \\
and Howard D. Trottier$^{(b)}$\cite{HDT}}

\vskip 12 pt

\begin{instit}
$^{(a)}$Physics Department, McGill University, 3600 University St., \\
Montreal, Quebec, Canada H3A 2T8
\end{instit}

\vskip 12 pt

\begin{instit}
$^{(b)}$TRIUMF, 4004 Wesbrook Mall, Vancouver, B.C., Canada V6T 2A3
\end{instit}

\begin{abstract}
We analyze the electromagnetic amplitude for the leptonic decays of
pseudoscalar mesons in the quark model, with particular emphasis on
$\eta \to l^+ l^-$ ($l=e,\mu$). We evaluate the electromagnetic box
diagram for a quark-antiquark pair with an arbitrary distribution of
relative three-momentum ${\bf p}$: the amplitude is obtained to
{\it all orders\/} in ${\bf p} / m_q$, where
$m_q$ is the quark mass. We compute
$B_P \equiv \Gamma(\eta \to l^+ l^-) /
\Gamma(\eta \to \gamma\gamma)$
using a harmonic oscillator wave
function that is widely used in nonrelativistic (NR) quark model
calculations, and with a relativistic momentum space wave function
that we derive from the MIT bag model. We also compare with a
quark model calculation in the limit of extreme NR binding due
to Bergstr\"om. Numerical calculations of $B_P$ using these
three parameterizations of the wave function agree to within a
few percent over a wide kinematical range. Our results show that
the quark model leads in a natural way to
a negligible value for the ratio of dispersive to absorptive
parts of the electromagnetic amplitude for
$\eta \to \mu^+ \mu^-$ (unitary bound). However we find
substantial deviations from the unitary bound in
other kinematical regions, such as $\eta,\pi^0 \to e^+ e^-$.
Using the experimental branching ratio for
$\eta \to \gamma\gamma$ as input, these quark models yield
$B(\eta \to \mu^+\mu^-) \approx 4.3 \times 10^{-6}$,
within errors of the recent SATURNE measurement of
$5.1 \pm 0.8 \times 10^{-6}$, and
$B(\eta \to e^+ e^-) \approx 6.3 \times 10^{-9}$.
While an application of constituent quark models to the
pion should be viewed with particular caution, the branching ratio
$B(\pi^0 \to e^+ e^-) \approx 1.0 \times 10^{-7}$
is independent of the details of the above quark model wave
functions to within a few percent.

\end{abstract}

%%% This command resulted in
%%% a blank page with only "Typset in REVTEX"
%%% \pacs{1990 PACS number(s): }
\newpage

\narrowtext

\section{\bf INTRODUCTION}
The rare leptonic decays of pseudoscalar mesons, such as
$\pi^0 \to e^+ e^-$ and $\eta,K \to l^+ l^-$ ($l=e,\mu$),
provide sensitive probes of new physics both in and
beyond the standard model \cite{Topical}. In the case of the
$\pi^0$ and $\eta$ decays, a careful analysis of the
electromagnetic contribution to the amplitude is required in
order to isolate a possible contribution from new physics.
While the absorptive part of the electromagnetic amplitude
can be reliably determined by unitarity from experimental data
on the two-photon width of the pseudoscalar, theoretical
estimates of the dispersive part are in some
disagreement \cite{etaReviews}.

An improved experimental measurement of the branching ratio
for $\eta \to \mu^+\mu^-$ has recently been obtained at
the $\eta$ meson facility SATURNE in Saclay \cite{SAT}, with
the result $B(\eta \to \mu^+\mu^-) = 5.1 \pm 0.8 \times 10^{-6}$.
This is to be compared with the lower limit obtained by
neglecting the dispersive part of the electromagnetic amplitude,
yielding the so-called unitary bound
$B(\eta \to \mu^+\mu^-) \geq B^{\rm unit} = 4.3 \times 10^{-6}$.

The SATURNE measurement is significantly closer to the
unitary limit than previous experiments (for a compilation
of earlier measurements see Ref. \cite{PDG}). It is therefore
of interest to reconsider the theoretical situation
with respect to the magnitude of the dispersive contribution
to the electromagnetic amplitude in this general class of decays.

The unitary bound for the leptonic decay of a pseudoscalar
meson ${\cal P}$ is most conveniently expressed in terms of
the ratio of leptonic to two-photon widths
\begin{equation}
   B_P \equiv
   { \Gamma({\cal P} \to l^+l^-) \over
     \Gamma({\cal P} \to \gamma\gamma) }
   \equiv {1 \over 2}
   \left( {\alpha m_l \over \pi m_P} \right)^2
   v \vert R \vert^2 ,
\label{BP}
\end{equation}
where $m_P$ is the meson mass,
and $v$ is the lepton velocity in the center-of-mass:
\begin{equation}
   v = \sqrt{ 1 - 4 {m_l^2 \over m_P^2} } .
\label{v}
\end{equation}

To establish our notation, we express $R$ in terms of
a ratio of the amplitudes for the two modes.
The invariant amplitude for the leptonic decay
can be parameterized as
\begin{equation}
   {\cal M}({\cal P} \to l^+l^-) \equiv
   {e^4 \over 16 \pi^2} m_l \,
   \bar u(\bar k) \gamma^5 v(k)  L ,
\label{Mll}
\end{equation}
where the explicit factor of the lepton mass $m_l$ reflects
the helicity flip that occurs over the lepton line in
Fig. \ref{FigLL}. The width is given by
\begin{equation}
   \Gamma({\cal P} \to l^+ l^-) = {\alpha^4 \over 8\pi}
   m_l^2 m_P v \vert L \vert^2 .
\label{Gll}
\end{equation}
The amplitude for the two-photon decay takes the form
\begin{equation}
   {\cal M}({\cal P} \to \gamma\gamma) \equiv
   i e^2 \epsilon_{\mu\nu\alpha\beta}
   \varepsilon_1^\mu \varepsilon_2^\nu
   q_1^\alpha q_2^\beta
   F ,
\label{Mgg}
\end{equation}
where $\varepsilon_{1,2}$ and $q_{1,2}$ are the polarizations and
momenta of the two photons. The form factor $F$ is purely real
for the decay to on-shell photons. The width is
\begin{equation}
   \Gamma({\cal P} \to \gamma \gamma) = {\alpha^2 \over 4}
   \pi m_P^3 F^2 ,
\label{Ggg}
\end{equation}
and the form factor $R$ for the relative branching ratio $B_P$ of
Eq. (\ref{BP}) follows
\begin{equation}
   R \equiv {L \over F} .
\label{RLF}
\end{equation}

Unitarity implies a connection between ${\rm Im} \, L$ and $F$,
resulting in the following model-independent result for
the absorptive part of $R$ \cite{Unitarity}
\begin{equation}
   {\rm Im} \, R = {\pi \over v}
   \ln\left( {1 - v \over 1 + v} \right) .
\label{ImR}
\end{equation}
The unitary bound on $B_P$ is obtained by assuming
that the dispersive part of $R$ is negligible. In the case
of $\eta \to \mu^+\mu^-$, Eq. (\ref{ImR}) implies
$B_P \geq B_P^{\rm unit} = 1.1 \times 10^{-5}$.
When combined with the experimental result for the two-photon
branching ratio (Ref. \cite{PDG}), this leads to the unitary
limit for $B(\eta \to \mu^+ \mu^-)$ quoted above.
In general, a nonvanishing dispersive part leads to
a correction to the unitary bound given by
\begin{equation}
   {B_P \over B_P^{\rm unit}} =
   1 + \left( { {\rm Re} \, R \over {\rm Im} \, R } \right)^2 .
\label{BBunit}
\end{equation}

Most theoretical calculations of the dispersive part
of $R$ have been based on a point-like coupling of the
pseudoscalar to off-shell photons \cite{Disp},
including dispersion relations and vector meson dominance.
Alternatively, a point-like coupling of the pseudoscalar to
nucleons or quarks (which then decay to virtual photons
through a triangle diagram) has also been considered \cite{etaNN}.
Recently, a calculation of $\eta$ and $\pi$ meson leptonic
decays has been made in chiral perturbation
theory \cite{Chiral}.

We consider instead a bound state description of the quarks
which comprise the meson. We evaluate the leptonic decay
${\cal P} \to l^+l^-$ at the quark level, which proceeds
through the one-loop diagrams illustrated in Fig. \ref{FigLL}.
This approach was considered by Bergstr\"om \cite{Berg} in
the limit of extreme nonrelativistic binding,
where the bound state quarks are assumed to be at rest
in the center of mass of the meson (and the meson mass
is assumed to be exactly twice the quark rest mass).
These approximations must be viewed with caution however
when applied to light mesons such as the $\eta$.

In this paper we allow the quarks to have an arbitrary
distribution of relative three-momentum ${\bf p}$. We compute
the quark model amplitude for ${\cal P} \to l^+ l^-$ to
{\it all orders\/} in ${\bf p} / m_q$, where $m_q$ is
the quark mass. A quark model wave function supplies
the distribution of relative three-momenta for the
quark-antiquark ($q \bar q$) bound state.

This approach is well known in nonrelativistic (NR) quark models
\cite{WVR,JR,HayneIsgur,MockMeson}, and has been used in a
variety of applications, including calculations of the two-photon
widths of light pseudoscalars \cite{BergGG,HayneIsgur}.
We make use of two parameterizations of the momentum space
wave function: a harmonic oscillator form that is widely used
in NR quark model calculations \cite{HayneIsgur,MockMeson},
and a new relativistic parameterization that we derive from
the MIT bag model. We also compare with the extreme
nonrelativistic limit obtained by Bergstr\"om.

Although the NR quark model and the MIT bag model
both give successful phenomenological descriptions of light
hadrons, we use these models here mainly because they
provide simple analytical expressions for the momentum space
wave function in two limits that characterize a wide class
of models. Our purpose in this paper is to make an
estimate of momentum-dependent effects in the quark model
amplitude for ${\cal P} \to l^+ l^-$, rather than to assess
the the predictive power of a particular model.

We expect that some of the model dependence inherent in
a description of light quark binding may cancel in the
relative branching fraction $B_P$. In fact, our
numerical calculations using the above parameterizations
of the wave function agree to within a
few percent over a wide kinematical range.
Our results show that the quark model leads in a
natural way to a negligible value for the ratio of
dispersive to absorptive parts of the electromagnetic amplitude
for $\eta \to \mu^+ \mu^-$. On the other hand, we find
substantial deviations from the unitary bound in other
kinematical regions, such as $\eta,\pi^0 \to e^+ e^-$.

The rest of this paper is organized as follows.
In Sec.~II, we describe our method for the evaluation
of ${\cal P} \to l^+ l^-$ for a quark-antiquark pair with an
arbitrary distribution of relative three-momentum. We evaluate the
electromagnetic box diagram for the leptonic decay in closed form.
In Sec.~III, we present quantitative
results using a harmonic oscillator wave function.
In Sec.~IV we derive a relativistic momentum space wave
function based on the MIT bag model, which we also use
to obtain quantitative results.
We summarize our findings in Sec.~V.

\section{\bf ``MOCK MESON'' METHOD}

\subsection{\bf General framework}

A conventional approach to the evaluation of hadronic
matrix elements is to decompose the bound state
into a superposition of free plane wave quark-antiquark
($q \bar q$) pairs \cite{WVR,JR,HayneIsgur}.
An economical and physically reasonable description of the momentum
space wave function is obtained by assuming that the
quark and antiquark have equal and opposite three-momentum
in each component of the plane wave expansion.
In this ``mock meson'' description,
the momentum-space state vector $\vert M(P) \rangle$ for
a meson in the center-of-mass frame [$P=(m_P,\vec 0)$]
has the following decomposition \cite{HayneIsgur}
\begin{equation}
   \vert M(P) \rangle \equiv
   \sqrt{2 m_P} \int { d^3{\bf p} \over (2\pi)^{3/2} }
   \Phi({\bf p}) {1 \over 2 E_p}
   \vert q({\bf p}) \rangle
   \vert \bar q (-{\bf p}) \rangle ,
\label{Mock}
\end{equation}
where $E_p \equiv \sqrt{ {\bf p}^2 + m_q^2 }$, and $m_q$
is the quark mass. We have omitted color, spin, and flavor indices
in Eq. (\ref{Mock}) for convenience.
We use invariant normalizations throughout this paper, e.g.,
$\langle q({\bf p}') \vert q({\bf p}) \rangle =
2 E_p (2\pi)^3 \delta^3({\bf p'} - {\bf p})$. The
wave packet amplitude $\Phi({\bf p})$
is normalized according to
\begin{equation}
   \int d^3{\bf p} \vert \Phi({\bf p}) \vert^2 = 1 .
\label{Phinorm}
\end{equation}
For the ground state pseudoscalar mesons of interest here,
we assume that the wave function is spherically symmetric,
$\Phi({\bf p}) = \Phi(p)$.

The right hand side of Eq. (\ref{Mock}) is a (zero) momentum
eigenstate, by construction. However, since the quarks are
taken to be on-shell, the energies of the individual
$q \bar q$ plane wave components are in general not equal
to the bound state energy. This energy ``mismatch'' leads
to several ambiguities in mock meson calculations,
including the value to be used for the total $q \bar q$
energy running through intermediate states in the amplitude,
and in phase space factors \cite{HayneIsgur,MockMeson}.
A popular prescription is to identify the total
$q \bar q$ energy appearing in the amplitude
with the mean energy of the wave packet \cite{HayneIsgur}.
In the case of quark model calculations of the pseudoscalar
two-photon width for example \cite{HayneIsgur}, this
prescription leads to a phase space dependence on the
meson mass that is in agreement with phenomenological
estimates of the $\eta-\eta'$ mixing angle
(see, e.g., Ref. \cite{PDG}).

For the purpose of calculating the amplitude
$R$ however, it is crucial to take the total energy
of the wave packet running through intermediate states
to be equal to the physical meson mass (the relative
branching fraction $B_P$ is very insensitive to
overall phase space factors, on the other hand).
This is necessary in order to obtain the correct unitarity relation
Eq. (\ref{ImR}) between the absorptive part of the
amplitude for the leptonic decay, and the (on-shell)
two-photon matrix element. This prescription has also
been used in other ``mock meson'' calculations \cite{CapGod}.

In the extreme nonrelativistic limit Eq. (\ref{Mock}) implies
that the $q \bar q$ annihilation amplitude is proportional to the
coordinate-space wave function (or its derivatives) at the
origin \cite{WVR,JR}. This limit has been used for example
to make calculations of heavy quarkonium matrix
elements \cite{Quarkonium}, and was used by Bergstr\"om
in a calculation of ${\cal P} \to l^+ l^-$ \cite{Berg}.

Equation (\ref{Mock}) has been used to {\it all orders\/}
in the $q \bar q$ relative momentum in nonrelativistic
quark model calculations of various matrix elements
\cite{HayneIsgur,MockMeson}, including pseudoscalar meson
decay to two photons, Eq. (\ref{Mgg}).
The form factor $F$ is found by evaluation of the Feynman
diagrams in Fig. \ref{FigGG} \cite{BergGG,HayneIsgur}
\begin{equation}
   F = { 4 \sqrt3 \over \sqrt{2\pi} }
   { m_q \over m_P^{3/2} } \, Q^2
   \int p dp \, \Phi(p)
   \ln \left[ {E_p + p \over E_p - p} \right] ,
\label{Fint}
\end{equation}
where the logarithm comes from an integration of
the quark propagator over angles (assuming that
the wave function $\Phi$ is spherically symmetric).
$Q$ is the quark charge in units of the proton charge.
For decays of flavor-mixed states such as the $\eta$, it
is understood that an expectation value of
the above expression for $F$ is taken between the
meson flavor wave function and the vacuum.

Equation (\ref{Fint}) is obtained by assuming that each
plane wave component of the $q \bar q$ wave packet
has the same total energy \cite{BergGG,HayneIsgur}. In our
case, this means that a factor of $1/m_P$ is extracted
from the quark propagator, which is contained in the
overall factor of the meson mass in Eq. (\ref{Fint}).

\subsection{\bf Application to leptonic decays}

We derive the mock meson amplitude for ${\cal P} \to l^+ l^-$
along the same lines which lead to Eq. (\ref{Fint})
for the two-photon matrix element:
\begin{equation}
   {\cal M}({\cal P} \to l^+ l^-) =
   \sqrt{2 m_P}
   \int { d^3 {\bf p} \over (2\pi)^{3/2} }
   {1 \over 2 E_p} \Phi(p)
   {\cal M}_{q \bar q} ({\bf p}) ,
\label{Md3p}
\end{equation}
where ${\cal M}_{q \bar q}$ is the amplitude for a given
three-momentum component of the $q \bar q$ wave packet
\begin{eqnarray}
   & & \quad\quad {\cal M}_{q \bar q} ({\bf p}) =
   -i 2 \sqrt3 e^4 Q^2
   \int {d^4 q \over (2\pi)^4} Q^{\mu\nu}(q) L_{\mu\nu}(q)
   \nonumber \\
   & & \times { 1 \over
   [ q^2 + i\epsilon ]
   [ (P - q)^2 + i\epsilon ]
   [ (p - q)^2 - m_q^2 + i\epsilon ]
   [ (q - k)^2 - m_l^2 + i\epsilon ]  } .
\label{Mqq}
\end{eqnarray}
The factor of 2 above accounts for the equal contribution of
the two diagrams in Fig. \ref{FigLL}, and $\sqrt3$ is a color factor.
$Q^{\mu\nu}$ is the spin-singlet $q \bar q$ current
\begin{equation}
   Q^{\mu\nu} \equiv
   {1 \over \sqrt2} \sum_{\lambda=\pm} \lambda
   \bar v_{-\lambda}(-{\bf p})
   \gamma^\nu \left[ \not\!p \, - \!\not\!q + m_q \right]
   \gamma^\mu u_\lambda({\bf p})
   = {4 i \over \sqrt2} m_q \epsilon^{0\alpha\mu\nu} q_\alpha ,
\label{Qpol}
\end{equation}
where, e.g., $u_\lambda({\bf p})$ is a positive energy spinor
of three-momentum ${\bf p}$ and angular momentum
$\lambda/2$ along a fixed quantization axis.
$L^{\mu\nu}$ is the spin-singlet projection of
the lepton current
\begin{equation}
   L^{\mu\nu} = 2 i {m_l \over m_P}
   \epsilon^{0\alpha\mu\nu} q_\alpha \,
   \bar u(\bar k) \gamma^5 v(k)
\label{Lpol}
\end{equation}
[spin-triplet components of $L^{\mu\nu}$
vanish under contraction with $Q^{\mu\nu}$, or under
integration in Eq. (\ref{Md3p})].

Comparison of Eqs. (\ref{Md3p})--(\ref{Lpol}) with
Eq. (\ref{Mll}) yields the following expression%%
\footnote{As with Eq. (\ref{Fint}) for the on-shell
two-photon form factor, it is understood that an
expectation value of Eq. (\ref{Lint}) for $L$ is taken
in the case of a flavor-mixed state.}
for the form factor $L$
\begin{equation}
   L = 16 \sqrt3 {m_q \over \sqrt{m_P}} Q^2
   \int { d^3 {\bf p} \over (2\pi)^{3/2} }
   \Phi(p) {1 \over E_p} I({\bf p}) ,
\label{Lint}
\end{equation}
where
\begin{equation}
   I({\bf p}) = {1 \over i\pi^2}
   \int d^4 q { - {\bf q}^2 \over
     [ q^2 + i\epsilon ]
     [ (P - q)^2 + i\epsilon ]
     [ (p - q)^2 - m_q^2 + i\epsilon ]
     [ (q - k)^2 - m_l^2 + i\epsilon ] } .
\label{Iloop}
\end{equation}
This integral can be evaluated analytically.
An identity relating ${\bf q}^2$ to
a linear combination of (inverse) propagators reduces
the integral to a sum of scalar three- and four-point
functions, plus an integral of three propagators with a factor
of $q^0$ in the numerator. A fictitious photon mass is
introduced in intermediate calculations as the scalar vertex
and box functions obtained in this way are infrared
divergent (the total integral is infrared
finite). We evaluate the divergent integrals using
expressions provided in Ref. \cite{ScalarInts}.
After some algebra, we find:
\begin{equation}
   I({\bf p}) =
   { 1 \over 4 \sqrt{ (p \cdot k)^2  -  m_q^2 m_l^2 } }
   \Bigl[ \ln(x_p) I_L + I_R \Bigr] ,
\label{Ip}
\end{equation}
where
\begin{equation}
   I_L \equiv
   2 + \ln\left( {m_q m_l \over m_P^2} \right)
   + 2 \ln(1 - x_p^2) - {1 \over 2} \ln(x_p)
   + i \pi ,
\label{IL}
\end{equation}
\begin{equation}
   I_R \equiv
   - {\pi^2 \over 6}
   + {1 \over 2} \ln^2 \left( {m_l \over m_q} \right)
   + {\rm Sp}(x_p^2)
   + {\rm Sp} \left( 1 - x_p {m_l \over m_q} \right)
   + {\rm Sp} \left( 1 - x_p {m_q \over m_l} \right) ,
\label{IR}
\end{equation}
and
\begin{equation}
   x_p \equiv
   {p \cdot k  -  \sqrt{ (p \cdot k)^2 - m_q^2 m_l^2 }
   \over  m_q m_l} .
\label{xp}
\end{equation}
${\rm Sp}(x)$ is the Spence function
\begin{equation}
   {\rm Sp}(x) = - \int_0^x dt {\ln(1 - t) \over t} .
\label{Spence}
\end{equation}

A crucial intermediate step in
our evaluation of the loop integral in Eq. (\ref{Iloop})
is the identification of the total energy of each component
of the $q \bar q$ wave packet
with the physical meson mass (i.e., $2 E_p \equiv m_P$).
This is necessary in order to satisfy the unitarity
relation of Eq. (\ref{ImR}). In particular, without
this prescription for the wave packet energy, the loop
integral would acquire unphysical branch cuts that do not
correspond to the ``unitarity cut'' through the
intermediate photons in Fig. \ref{FigLL}.
On the other hand, we do use the actual plane wave
energy $p^0 = E_p = \sqrt{ {\bf p^2} + m_q^2 }$
in our final expression Eq. (\ref{Ip}) for
$I({\bf p})$. We note that a similar prescription has been
used in other mock meson calculations, such as the two-photon
width (cf. Eq. (\ref{Fint}) and Refs. \cite{BergGG,HayneIsgur}).

The fact that our final expression for ${\rm Im} \, L$
exactly satisfies the unitarity relation of Eq. (\ref{ImR})
is a nontrivial check of the above prescription for handling
the ambiguity in the wave packet energy. Indeed, after an
analytical evaluation of the angular integration
over ${\rm Im} \, I({\bf p})$
[Eqs. (\ref{Lint}), (\ref{Ip}), and (\ref{IL})],
we find that ${\rm Im} \, L$
can be written as a product of Eq. (\ref{Fint}) for the
on-shell two-photon form factor $F$, times the
right hand side of Eq. (\ref{ImR}).

\section{\bf HARMONIC OSCILLATOR WAVE FUNCTION}

We evaluate the quark model form factor $R$
[Eqs. (\ref{BP}), (\ref{RLF}), (\ref{ImR}),
(\ref{Fint}) and (\ref{Lint})] using a harmonic oscillator
parameterization of the $q \bar q$ momentum space
wave function that has been widely used in nonrelativistic
quark model calculations \cite{HayneIsgur,MockMeson}
\begin{equation}
   \Phi_{\rm osc}({\bf p}) \equiv (\beta^2 \pi)^{-3/4}
   \exp\left( -{ {\bf p}^2 \over 2\beta^2} \right) .
\label{PhiOsc}
\end{equation}
Some representative results for the ratio
${\rm Re} \, R  /  {\rm Im} \, R$ obtained with this
wave function are given in Fig. \ref{FigROsc}.
For the sake of illustration, in this figure we
take the ``physical'' mass of the hypothetical
meson to be equal to the mean energy of the wave packet
($m_P = 2 \langle E_p \rangle$).

In the limit $\beta / m_q \to 0$, the wave packet
becomes nonrelativistic, with $\langle E_p \rangle \to m_q$.
Bergstr\"om has made a quark model calculation of $R$
in the extreme nonrelativistic (ENR) limit, taking the
quark and antiquark to be at rest in the center of mass
of the meson, with $2 m_q \equiv m_P$ \cite{Berg}.
He found:
\begin{equation}
   {\rm Re} \left( R_{\rm ENR} \right) = {2 \over v}
   \left\{
   {1 \over 4} \ln^2 \left( {1 + v \over 1 - v} \right)
   - \ln \left( {1 + v \over 1 - v} \right)
   + {\pi^2 \over 12}
   + {\rm Sp} \left( - {1 - v \over 1 + v} \right)
   \right\} .
\label{BergReR}
\end{equation}
Our results agree with Eq. (\ref{BergReR})
in the limit $\beta / m_q \to 0$.

We see that the real part of $R$ is generally small compared
to its imaginary part, except for small lepton masses
$m_l \ll m_P$, or in the limit of ultrarelativistic binding
$m_q \ll m_P$ (${\rm Re} \, R$ exhibits a
logarithmic divergence as $m_l$ or $m_q$ go to zero, which is
characteristic of mass singularities in loop processes).%%
\footnote{We include curves in Fig. \ref{FigROsc}
to illustrate the results of a harmonic oscillator calculation
in the ultrarelativistic region; however, the physical region
in this model corresponds to $\beta / m_q$ of order unity
or smaller.}
Since the dispersive part of $R$ appears
quadratically in the branching ratio, we can expect
small corrections to the unitary limit over a
wide kinematical range.

To make contact with experimental data on $\eta \to l^+ l^-$,
we assign physical values to the quark model parameters
and meson mass. Various sets of values for the quark masses
and the wave function Gaussian parameter $\beta$ have been used
in the literature \cite{HayneIsgur,MockMeson}.
Our results for the relative branching fraction $B_P$
are insensitive to variations in these parameters
over a wide range.%%
\footnote{The absolute normalizations of the
individual widths $\Gamma({\cal P} \to \gamma\gamma)$
and $\Gamma({\cal P} \to \mu^+ \mu^-)$ depend strongly
on the choice of model parameters. The widths also
depend on the choice of total $q \bar q$ energy that
is used in the phase space \cite{HayneIsgur,MockMeson}.
The oscillator wave function results for the
two-photon widths of the $\pi^0$, $\eta$ and $\eta'$
are generally within a factor of a few of the
experimental values \cite{HayneIsgur}.}
We use the typical values \cite{OscParms}
\begin{eqnarray}
   & & m_u = m_d = 330~{\rm MeV} ,  \nonumber \\
   & & m_s = 550~{\rm MeV} ,  \\
   & & \beta = 310~{\rm MeV} ,  \nonumber
\label{OscVals}
\end{eqnarray}
and an $\eta-\eta'$ mixing angle of
$-20^\circ$ (cf. Ref. \cite{PDG}).

We plot the relative branching fraction $B_P$
compared to the unitary bound $B_P^{\rm unit}$
[Eq. (\ref{BBunit})] in Fig. \ref{FigBP}, for a
range of lepton masses. We take $m_P$ equal
to the physical $\eta$ mass when evaluating
Eqs. (\ref{Fint}) and (\ref{Lint}). We also include a
plot of $B_P$ as obtained by Bergstr\"om in the limit of extreme
nonrelativistic binding (cf. Eq. (\ref{BergReR})
and Ref. \cite{Berg}). The harmonic oscillator
and ENR results differ by less than 5\% over the
entire kinematical region in $m_l / m_\eta$.

Using the experimental value for the two-photon branching
fraction as input \cite{PDG}, we find
$B(\eta \to \mu^+ \mu^-) = 4.3 \times 10^{-6}$ and
$B(\eta \to e^+ e^-) = 6.3 \times 10^{-9}$
in the harmonic oscillator model.
The result for the decay to muons is within errors
of the recent SATURNE measurement
$B_{\rm exp}(\eta \to \mu^+ \mu^-) = 5.1 \pm 0.8 \times 10^{-6}$
\cite{SAT}, and is only 0.2\% larger than the unitary limit.
On the other hand, the branching fraction to electrons is
$\approx 3.6$ times larger than the corresponding unitary limit.
We also compute $B(\pi^0 \to e^+ e^-) = 1.0 \times 10^{-7}$
in the oscillator model (about twice the unitary limit),
although an application of constituent quark models to the
pion should be viewed with particular caution.

\section{\bf BAG MODEL-INSPIRED RELATIVISTIC WAVE FUNCTION}

Although the mock meson method described in Sec.~IIA
was developed in the context of nonrelativistic quark models
\cite{HayneIsgur,MockMeson}, we find that the
MIT bag model \cite{MIT}
can be used to motivate a relativistic parameterization
of Eq. (\ref{Mock}). We note in this
connection that the momentum space wave packet distribution
$\Phi({\bf p})$ is, in principle, an arbitrary function,
and need not be localized around momenta that are
small compared to the quark mass \cite{HayneIsgur}.

To begin with, we must take account of the fact that
bound states in the usual bag model are not momentum eigenstates,
while the quark and antiquark in Eq. (\ref{Mock}) for the
mock meson have equal and opposite three-momentum
by construction. Our approach is complementary to the formalism
introduced by Donoghue and Johnson, who introduced a wave packet
in order to decompose the bag model wave function into momentum
eigenstates \cite{DJ}. Their wave packet is determined by
normalizing to a particular matrix element, such as $f_\pi$.
We choose instead to identify the wave packet amplitude
$\Phi({\bf p})$ in Eq. (\ref{Mock}) directly from the
Fourier transform of the cavity wave function.

To begin with, consider the wave function for the ground state
of a single quark in a cavity of radius $R$
\begin{equation}
   \psi_\lambda ({\bf r}) =
   {\cal N} \left[
   \begin{array}{c}
       \left( {\omega_0 + m_0 \over \omega_0} \right)^{1/2}
       i j_0 \left( {x r \over R} \right)  U_\lambda \\
     - \left( {\omega_0 - m_0 \over \omega_0} \right)^{1/2}
         j_1 \left( {x r \over R} \right)
       \mbox{\boldmath$\sigma$} \! \cdot \! {\bf\hat r} \, U_\lambda
   \end{array}
   \right] \quad (r \leq R) ,
\label{WF}
\end{equation}
where for later use we label the ``current'' quark mass
by $m_0$, and where $\omega_0 \equiv (x^2/R^2 + m_0^2)^{1/2}$.
The momentum eigenvalue $x$ is determined by the boundary
condition $\bar\psi\psi\vert_{r=R}=0$ \cite{MIT}.
${\cal N}$ is a normalization, and $U_\lambda$ is a
two-component spinor with polarization $\lambda = \pm$ along
the $z$ axis.

The Fourier transform of a Dirac wave function can be
found by standard methods \cite{BjDr}.
Since we can decompose the wave function along an
arbitrary complete set of states, the plane wave spinors
in the Fourier transform of Eq. (\ref{WF})
need not have the same (``current'') quark mass $m_0$
as $\psi_\lambda({\bf r})$. We compute a
Fourier transform along plane waves
of arbitrary ``effective'' mass $m_{\rm eff}$
\begin{equation}
   \psi_\lambda({\bf r}) =
   \int { d^3{\bf p} \over (2 \pi)^{3/2} }
   { 1 \over \sqrt{2 \omega_{\rm eff}(p)} }
   \left[
   \phi(p)
   u_\lambda({\bf p})
   e^{ i {\bf p} \cdot {\bf r} }
   +
   \widetilde\phi(p)
   \sum_{\lambda'=\pm} S_{\lambda\lambda'}({\bf\hat p})
   v_{\lambda'}({\bf p})
   e^{ -i {\bf p} \cdot {\bf r} }
   \right] ,
\label{Pspace}
\end{equation}
where $u_\lambda({\bf p})$ and $v_\lambda({\bf p})$ are
plane wave Dirac spinors with angular momentum $\lambda/2$
along the $z$ axis, and
$S_{\lambda\lambda'}({\bf\hat p}) \equiv U_\lambda^\dagger
\mbox{\boldmath$\sigma$} \! \cdot \! {\bf\hat p} \, U_{\lambda'}$.
The spinors are normalized to
$u^\dagger u = - v^\dagger v = 2 \omega_{\rm eff}(p)$, where
$\omega^2_{\rm eff}(p) \equiv {\bf p}^2 + m_{\rm eff}^2$
\cite{BagEave}.

The Fourier amplitudes $\phi$ and $\widetilde\phi$
depend on the magnitude of the three momentum
$p \equiv \vert {\bf p} \vert$. We find \cite{BR}
\begin{eqnarray}
   \phi(p) & = &
   { 1 \over \sqrt{2 \pi} }
   \left[ \psi^\dagger(0) \psi(0) \right]^{1/2}
   {R^2 \over p x}
   \left( { \omega_{\rm eff}(p) + m_{\rm eff}  \over
          2 \omega_{\rm eff}(p) } \right)^{1/2}
\nonumber \\
   & \times &
   \Biggl\{ j_0(pR - x) - j_0(pR + x)
   + \left( {\omega_0 - m_0 \over \omega_0 + m_0} \right)^{1/2}
\nonumber \\
   & & \qquad \times
   { p \over \omega_{\rm eff}(p) + m_{\rm eff} }
   \left[ j_0(pR + x) + j_0(pR - x) - 2 j_0(pR) j_0(x) \right]
   \Biggr\} ,
\label{phi}
\end{eqnarray}
and
\begin{eqnarray}
   \widetilde \phi(p) & = &
   { 1 \over \sqrt{2 \pi} }
   \left[ \psi^\dagger(0) \psi(0) \right]^{1/2}
   {R^2 \over p x}
   \left( { \omega_{\rm eff}(p) + m_{\rm eff} \over
          2 \omega_{\rm eff}(p) } \right)^{1/2}
\nonumber \\
   & \times & \Biggl\{
   - { p \over \omega_{\rm eff}(p) + m_{\rm eff} }
   \left[ j_0(pR - x) - j_0(pR + x) \right]
   + \left( {\omega_0 - m_0 \over \omega_0 + m_0} \right)^{1/2}
\nonumber \\
   & & \qquad \times
   \left[ j_0(pR + x) + j_0(pR - x) - 2 j_0(pR) j_0(x) \right]
   \Biggr\} .
\label{phit}
\end{eqnarray}

The usual mock meson approach to the calculation of hadronic
matrix elements assumes that the quarks propagate as
free particles in intermediate states \cite{HayneIsgur}.
For example, the explicit factor of the quark mass in the
matrix elements for ${\cal P} \to \gamma\gamma, l^+ l^-$
[Eqs. (\ref{Fint}) and (\ref{Lint})]
is due to the helicity flip along an intermediate
quark line. In typical nonrelativistic
quark model calculations, the quarks have large
``constituent'' masses. This means, for example,
that the $u$, $d$ and $s$ quark components of the
$\eta$ and $\eta'$ flavor wave functions make comparable
contributions to their matrix elements in these
models, in agreement with simple phenomenological estimates
of the pseudoscalar mixing angle (see, e.g., Ref. \cite{PDG}).

In typical MIT bag model calculations however the quarks
are assigned current masses, in particular, $m_{u,d} \approx 0$.
In order to obtain a sensible phenomenological description of
helicity-flip amplitudes in a mock meson approach,
we must assume that a cavity quark propagates with
a mass that is different from its ``bare'' value $m_0$.
We therefore assign a constituent (or effective) quark
mass to intermediate quark lines, and for consistency
we use the same effective mass in the Fourier
transform Eq. (\ref{Pspace}) of the cavity wave function.

A rigorous calculation of matrix elements in the context
of the bag model would use a cavity propagator for
intermediate quark lines \cite{BagProp}. This was in fact
done in a calculation of $\pi^0 \to \gamma\gamma$ in
Ref. \cite{Bagpi0}. We note that the cavity quark
propagator differs from the free propagator by some terms that
act, to some extent, like a (momentum-dependent) effective
mass. A calculation of the leptonic decay using cavity propagators
would be quite involved however. On the other hand,
our extension of the mock meson method to the bag model
permits a straightforward evaluation of the matrix element
incorporating much of the basic physics underlying
cavity perturbation theory [this is further illustrated
by our comments below Eq. (\ref{Phiphi})].

Although the actual value of the effective mass for ``free'' quark
propagators is somewhat ad hoc, this approach can
nevertheless can be used with justification in the
calculation of the form factor $R$ of interest in this
paper [Eq. (\ref{BP})]. This is due to the fact
that the {\it ratio\/} between the leptonic and two-photon
amplitudes is insensitive to variations in the
actual value of the effective quark mass over a
large range. Observe in particular that the same factor of the
intermediate quark mass due to the helicity flip appears
in both matrix elements [cf. Eqs. (\ref{Fint}) and (\ref{Lint})].
On physical grounds, the effective quark mass should
be on the order of the cavity energy $\omega_0$;
we find that $B_P$ changes by only a few percent as
$m_{\rm eff}$ is varied from ${\case1/4}\omega_0$ to $\omega_0$.
We use $m_{\rm eff} = \omega_0$ in the following calculations.

At any rate, we do not regard our results as providing
precise tests of the predictive power of the bag model.
Our interest in this paper is to make an estimate of
momentum-dependent effects in the correction to
the unitary limit for ${\cal P} \to l^+ l^-$ in the context
of general quark models. In that respect, the
bag model-inspired wave packet amplitudes provide
a very useful comparison with the harmonic oscillator
wave function used in Sec.~III. In particular,
$\phi(p)$ falls off only as $1/p^2$ at large $p$, compared
to the exponential decay of the oscillator wave function.

In this connection, we note that the mock meson
wave packet of Eq. (\ref{Mock}) is based on a single
particle description of the quark and the antiquark in the
bound state, which is in general inadequate to describe
localized relativistic states. This problem is shared
by the single wave packet amplitude introduced by
Donoghue and Johnson.
A proper connection between the localized cavity
wave function and momentum space eigenstates requires
the use of a Bogoliubov transformation \cite{BogT},
which in this case involves a mixing among the complete
set of cavity eigenstates.

On the other hand, we find that the negative energy components
of the Fourier transform actually make a small contribution
to the normalization of the ground state cavity mode.
The cavity wave function Eq. (\ref{WF}) is normalized to
one in the sphere of radius $R$, which implies
\begin{equation}
   \int d^3{\bf p}
   \left( \phi^2(p) + \widetilde\phi^2(p) \right) = 1 .
\label{phinorm}
\end{equation}
We find that the contribution of the negative energy component
to this normalization integral is at most 7\% in the
``ultrarelativistic'' limit ($m_0 R=0$), and falls
below 4\% by $m_0 R \sim 1$.
We conclude that a reasonable approximation to
the ground state cavity wave function can be obtained by
truncating the negative energy component of the
Fourier transform in Eq. (\ref{Pspace}),%%
\footnote{Note that the normalization of the wave packet
cancels in a calculation of the relative branching fraction $B_P$.}
and we therefore identify $\Phi({\bf p})$ in the
mock meson wave packet Eq. (\ref{Mock}) as
\begin{equation}
   \Phi_{\rm bag}({\bf p}) \equiv \phi(p)
\label{Phiphi}
\end{equation}
[cf. Eqs. (\ref{Phinorm}) and (\ref{phinorm})].

We note that ${\rm Re} \, R$ is very insensitive to the value
of the current quark mass $m_0$.
This is due in large measure to the fact that we use
an effective mass $m_{\rm eff} = O(\omega_0)$
for intermediate quark propagators. Thus, even for
a massless current quark, the effective mass is nonzero
($\omega_0 R \in [2.04,\infty]$ for $m_0 R \in [0,\infty]$).
The logarithmic mass singularity that would be
present in ${\rm Re} \, R$ for a truly massless
propagator is thereby avoided.
A similar situation occurs in a rigorous calculation
of the bag model width for $\pi^0 \to \gamma\gamma$
using cavity propagators for the massless
$u$ and $d$ quarks \cite{Bagpi0}.

We include our results for the relative branching
fraction $B_P(\eta \to l^+ l^-)$ for the physical
$\eta$ meson using this bag model-inspired
wave function in Fig. \ref{FigBP}. We use the
typical bag model parameter values \cite{MIT,etaBag}
\begin{eqnarray}
   & & m_u = m_d = 0 ,  \nonumber \\
   & & m_s = 300~{\rm MeV} ,  \\
   & & R = 3.3~{\rm GeV} ^{-1} . \nonumber
\label{BagVals}
\end{eqnarray}

The results for $B_P$ in the relativistic wave function
agree with the harmonic oscillator calculation (and
with the extreme nonrelativistic limit) to within
a few percent over wide range in $m_l / m_\eta$,
except for large lepton masses, near threshold.
This is evidently related to the fact that bag
model wave packet has appreciable components
at large $q \bar q$ relative momentum
[cf. $\Phi_{\rm bag}(p) \sim 1/p^2$ at large $p$,
compared to $\Phi_{\rm osc}(p) \sim \exp(-p^2)$].

For the physical cases of $\eta$ or $\pi^0$ decays to
muons or electrons, $m_l / m_P$ is small, and the
bag model-inspired results
$B(\eta \to \mu^+ \mu^-) = 4.4 \times 10^{-6}$,
$B(\eta \to e^+ e^-) = 6.1 \times 10^{-9}$, and
$B(\pi^0 \to e^+ e^-) = 1.0 \times 10^{-7}$
agree with the oscillator model to better than 5\%.

\section{\bf SUMMARY}

We have evaluated the electromagnetic box diagram for
the leptonic decay of a pseudoscalar quark-antiquark
pair with an arbitrary distribution of relative three-momentum.
Quantitative results were obtained in three different
models of the bound state wave function (a nonrelativistic
harmonic oscillator model \cite{HayneIsgur,MockMeson},
a new relativistic momentum space wave function that we derived from
the MIT bag model, and in the limit of extreme nonrelativistic
binding, analyzed previously by Bergstr\"om \cite{Berg}). Our
results demonstrate that the relative branching fraction
$B_P \equiv
B({\cal P} \to l^+ l^-) / B({\cal P} \to \gamma\gamma)$
is insensitive to the details of the
quark model wave function. In the case of $\eta$ and $\pi^0$
decays, the results obtained with the harmonic oscillator
and relativistic wave functions agree to within 5\%.

We find that the quark model leads in a natural way to
a negligible value for the ratio of dispersive to absorptive
parts of the electromagnetic amplitude for $\eta \to \mu^+ \mu^-$.
On the other hand, we find substantial deviations from the unitary
bound in other kinematical regions, such as $\eta \to e^+ e^-$.

Using the experimental branching ratio for $\eta \to \gamma\gamma$
as input, these quark models yield
$B(\eta \to \mu^+ \mu^-) \approx 4.3 \times 10^{-6}$,
within errors of the recent SATURNE measurement
of $5.1 \pm 0.8 \times 10^{-6}$, and
$B(\eta \to e^+ e^-) \approx 6.3 \times 10^{-9}$.
While an application of constituent
quark models to the pion should be viewed with particular
caution, the quark models considered here yield
$B(\pi^0 \to e^+ e^-) \approx 1.0 \times 10^{-7}$,
independent of the details of the model wave function to
within a few percent. This is to be compared with recent
experimental data from Fermilab,
$B_{\rm exp}(\pi^0 \to e^+ e^-) = 6.9 \pm 2.8 \times 10^{-8}$
\cite{FNAL}, and preliminary data from Brookhaven
$B_{\rm exp}(\pi^0 \to e^+ e^-) = 6.0 \pm 1.8 \times 10^{-8}$
\cite{BNL}.
The quark model branching ratios obtained here are comparable to
the results of a recent analysis in chiral perturbation
theory \cite{Chiral}, which requires the experimental value of
the branching ratio for $\eta \to \mu^+ \mu^-$ as input.

\acknowledgments
This work was supported in part by the Natural Sciences
and Engineering Research Council of Canada.

\figure{Quark model electromagnetic box diagrams for
${\cal P} \to l^+ l^-$. In a ``mock meson'' description (Sec. II),
the quark momenta in the center-of-mass are anti-correlated,
$p,\bar p = (p^0,\pm{\bf p})$. The lepton momenta are
$k,\bar k = (k^0,\pm{\bf k})$.\label{FigLL}}

\figure{Quark model diagrams for
${\cal P} \to \gamma\gamma$.\label{FigGG}}

\figure{Real part of the form factor $R$ for the
relative branching fraction $B_P({\cal P} \to l^+ l^-)$,
compared to the unitarity result for the imaginary part
[cf. Eqs. (\ref{BP}) and (\ref{ImR})],
in a harmonic oscillator model of the
bound state wave function [Eq. (\ref{PhiOsc})].
For this figure, the mass of the hypothetical meson
is set equal to the mean energy of the
$q \bar q$ wave packet.\label{FigROsc}}

\figure{Relative branching fraction $B_P$ for the leptonic
decay of the physical $\eta$ meson, compared
to the unitary bound $B_P^{\rm unit}$.
The three curves correspond a harmonic oscillator model
[Eqs. (\ref{PhiOsc}) and (\ref{OscVals})],
a bag model-inspired relativistic wave function
[Eqs. (\ref{Phiphi}) and (\ref{BagVals})], and
the limit of extreme nonrelativistic binding
[Eq. (\ref{BergReR})].\label{FigBP}}


\begin{references}
\bibitem{BagEave}We verify by numerical integration that the
expectation value of the ``current'' quark Hamiltonian operator
  $\mbox{\boldmath$\alpha$} \cdot
(-i\mbox{\boldmath$\nabla$}) + \beta m_0$
in the wave packet of Eq. (\ref{Pspace}) is equal to
the cavity eigenfrequency $\omega_0$ for
arbitrary effective quark mass $m_{\rm eff}$.
\end{references}
\end{document}